\documentclass[preprint,pra,aps,showpacs,preprintnumbers,amsmath,amssymb,eqsecnum]{revtex4}

\usepackage{graphicx}
\usepackage{dcolumn}
\usepackage{bm}
\usepackage{verbatim}
\usepackage{epsfig}
\usepackage{epstopdf}

\newcommand\beq{\begin{equation}}
\newcommand\eeq{\end{equation}}
\newcommand\bea{\begin{eqnarray}}
\newcommand\eea{\end{eqnarray}}

\def\a{{\bf a}}
\def\sgn{{\rm sgn}}

\def\ria{\rightarrow}

\def\s{{s}}

\def\half{\frac {1} {2}}

\def\x0{{{\bf x}_0}}

\begin{document}


\title{Two Proofs of Fine's Theorem}

\author{J.J.Halliwell}%
\email{j.halliwell@imperial.ac.uk}

\affiliation{Blackett Laboratory \\ Imperial College \\ London SW7
2BZ \\ UK }



\begin{abstract}
Fine's theorem concerns the question of determining the conditions under which a certain set of probabilities for pairs of four bivalent quantities may be taken to be the marginals of an underlying probability distribution. The eight CHSH inequalities are well-known to be necessary conditions, but Fine's theorem is the striking result that they are also a sufficient condition.
It has application to the question of finding a local hidden variables theory for measurements of pairs of spins for a system in an EPRB state.
Here we present two simple and self-contained proofs of Fine's theorem in which
the origins of this non-obvious result can be easily seen.
The first is a physically motivated proof which simply notes that this matching problem is solved using a local hidden variables model given by Peres.
The second is a straightforward algebraic proof which uses a representation of the probabilities in terms of correlation functions and takes advantage of certain simplifications naturally arising in that representation. A third, unsuccessful attempt at a proof, involving the maximum entropy technique is also briefly described.

\end{abstract}



\maketitle

\section{Introduction}

Consider the following simple but non-trivial problem in probability theory. We suppose
we are given a system described by four variables $\s_1, \s_2, \s_3, \s_4$ which may take values $\pm 1$, and which for convenience we call spins.
We suppose also that we are given the pair probabilities,
$ p( \s_1, \s_3)$, $ p (\s_1, \s_4)$, $p (\s_2, \s_3)$ and $ p (\s_2, \s_4) $.
Under what conditions are these pair probabilities the marginals of an underlying
probability for all four variables, $ p(\s_1, \s_2, \s_3, \s_4)$?

This question is of course very closely linked to the Clauser-Horne-Shimony-Holt (CHSH) analysis of an entangled
pair of spin states \cite{CHSH} and in this connection it is well-known that a necessary set of conditions is the eight CHSH inequalities,
\bea
-2 & \le & C_{13} + C_{14} + C_{23} - C_{24} \le 2
\label{CHSH1}
\\
-2 & \le & C_{13} + C_{14} - C_{23} + C_{24} \le 2
\label{CHSH2}
\\
-2 & \le & C_{13} - C_{14} + C_{23} + C_{24} \le 2
\label{CHSH3}
\\
-2 & \le & - C_{13} + C_{14} + C_{23} + C_{24} \le 2
\label{CHSH4}
\eea
where $C_{13}, C_{14}, C_{23}, C_{24}$ denote the correlation functions
\beq
C_{ij} = \sum_{s_1 s_2 s_3 s_4} s_i s_j \ p(\s_1,\s_2, \s_3, \s_4)
\label{Cij}
\eeq
and
\beq
\sum_{s_1 s_2 s_3 s_4}  \ p(\s_1,\s_2, \s_3, \s_4) =1
\eeq
The CHSH inequalities are easily derived by assuming that a probability $p(\s_1,\s_2, \s_3, \s_4) $ exists and then summing
it with simple inequalities of the form
\beq
-2 \le s_1 s_3 + s_1 s_4 + s_2 s_3 - s_2 s_4 \le 2
\label{ss}
\eeq
plus three more similar ones, thereby obtaining Eqs.(\ref{CHSH1})-(\ref{CHSH4}).

However, an important result due to Fine is that the CHSH inequalities are also a
{\it sufficient} condition for the existence of a probability matching the given marginals.
This intriguing result goes by the name of Fine's theorem.
Its proof is not as immediate or obvious as the proof of necessity. Fine gave a direct proof in Refs.\cite{Fine1,Fine2}
by showing by purely algebraic means how to flesh out the given set of marginals
into a full probability distribution. (A simple proof of the much easier problem
involving three bivalent variables, involving Bell's original inequalities
\cite{Bell}, was given by Suppes and Zanotti \cite{SuZa}.)
Pitowski \cite{Pit} gave a very different proof using the geometry of polytopes.
Garg and Mermin considered a general class of problems of
this type \cite{GaMer}, using properties of convex sets, and gave a proof of Fine's
theorem as an example. Generalizations of these ideas to $N$ qubits have been considered
by Zukowski and Brukner \cite{ZuBr}.

Since this result is far from obvious, it is of interest to find alternative proofs
which are clearer and more immediate.
The purpose of the present paper is therefore to give two self-contained proofs of Fine's theorem which are different and perhaps simpler than those cited above.
The idea is not to give a general solution to this matching problem, but
to give simple pedagogical pictures in which it is not hard to see
why the theorem is true.

The first proof is a physically-motivated one involving an explicit local hidden variables model
given by Peres \cite{Per2}. Clearly if a local hidden variables theory
exists matching the given marginals, then an underlying probability $p(s_1,s_2,s_3,s_4)$ exists so the CHSH inequalities must
be satisfied. The point here is to show that this statement is logically reversible for this
model -- if the CHSH inequalities are satisfied then the parameters of the local hidden variables
model may be chosen to match the given marginals so the sought-after probability solving the matching problem is that supplied by the local hidden variables theory.
In essence, we make a strategic guess as to the form of the underlying probability
and confirm that it solves the problem.

The second proof is a direct algebraic one, which takes advantage of
a particularly useful representation of the underlying probability in terms of its correlation functions,
\beq
p (\s_1, \s_2, \s_3, \s_4) = \frac {1} {16} \left( 1 +
\sum_i B_i s_i + \sum_{i<j} C_{ij} s_i s_j + \sum_{i<j<k} D_{ijk} s_i s_j s_k
+ E s_1 s_2 s_3 s_4 \right)
\label{ps}
\eeq
where the indices $i,j,k$ run over the values $1,2,3,4$ \cite{HaYe}. The correlation functions
$C_{ij}$ are given by Eq.(\ref{Cij}) and the remaining correlators
are given by
\bea
B_i &=& \sum_{\s_1 \s_2 \s_3 \s_4} \ \s_i \ p (\s_1, \s_2, \s_3, \s_4 )
\nonumber \\
D_{ijk} &=& \sum_{\s_1 \s_2 \s_3 \s_4} \ \s_i \s_j \s_k\  p (\s_1, \s_2, \s_3, \s_4)
\nonumber \\
E &=& \sum_{\s_1 \s_2 \s_3 \s_4}  \ \s_1 \s_2 \s_3 \s_4 \ p (\s_1, \s_2, \s_3, \s_4)
\eea
The marginals are then easily constructed by summing out some of the $\s_i$'s. So
for example
\beq
p(\s_1, \s_3) = \frac{1} {4} \left( 1 + B_1 \s_1 + B_3 \s_3 + C_{13} \s_1 \s_3 \right)
\label{p2}
\eeq
Note that we are using the mathematically incorrect but commonly employed notation in which
functions, such as $ p(s_1,s_3)$ are identified by their arguments. Also, we are assuming that the
single spin probabilities are consistent with the specified two-spin probabilities, so for example,
we assume that
\beq
\sum_{s_1} p( s_1, s_3) = p(s_3) = \sum_{s_2} p ( s_2, s_3)
\eeq

Eq.(\ref{p2}), plus three similar relations, mean that fixing the given four marginals is equivalent to fixing the values of $B_i$ for $i=1,2,3,4$
and the values of the four correlation functions $C_{13}$, $C_{23}$, $C_{14}$ and $C_{24}$.  The question of finding a probability matching the given marginals is then the question of whether the remaining unfixed correlation functions, $ C_{12}$, $C_{34} $, $D_{ijk}$ and $E$ can be chosen in such a way that the probability Eq.(\ref{ps}) is positive. However, as stated, we are not looking for the most general solution
to the problem, but instead seeking to show that {\it some} solution exists as long
as the CHSH inequalities are satisfied. This allows us to make a number of simplifications,
based on symmetries of the CHSH inequalities, as we shall see, and the algebraic solution
then turns out to be very straightforward.

We begin in Section 2 by briefly describing the related quantum problem from which
this question arises and we show from this how to argue that we may set the average spins, $B_i$, to zero. In Section 3 we describe the proof of Fine's theorem using a local hidden variable model.

Turning to the second algebraic proof, in Section 4 we solve algebraically a simpler problem
involving three variables, and in this case the necessary and sufficient conditions
are the four Bell inequalities. We give the algebraic solution to the main problem,
finding the conditions under which Eq.(\ref{ps}) is positive, in Section 5.
We summarize and conclude in Section 6. We also briefly describe a third attempt at proving Fine's theorem using an ansatz for the probability supplied by the maximum entropy technique, but this turns out to be unsuccessful.

This work arose directly from an early work about the use and misuse of
quasi-probabilities and their relation to Fine's theorem \cite{HaYe}. In particular,
the formula Eq.(\ref{ps}) was introduced there in the context of quasi-probabilities but has found particular use here as a genuine probability. This formula was also
written down earlier by Klyshko \cite{Kly}, who showed that a number of different
problems involving quantum ``paradoxes'' can reduce to a problem in probability theory
of matching given marginals. Fine's theorem appears to have had very wide impact
and many applications,
with his original paper receiving a very large number of citations, far too many
to discuss here in any detail. However, it is clearly very relevant to the questions
concerning the existence and interpretation of hidden variable theories
(see, for example Refs.\cite{SRBB,But}) and to generalizations of quantum
theory \cite{CDHMRS}. It may also have some role in the Leggett-Garg (or ``temporal
Bell'') inequalities \cite{LeGa,Ye1}, since they have the same form as the CHSH
inequalities, but this does not seem to have been explored.

\section{The Quantum Problem and a Simplification}

Some background and insight into the Fine problem may be obtained by considering
some aspects of the quantum-mechanical problem from which it arose.
The situation is the standard EPRB set up, in which
we consider a pair of particles $A$ and $B$ whose spins are in an entangled state.
(For general reviews of the Bell and CHSH inequalities in this area see
for example Refs.\cite{Gis,WeWo}).
The most famous example is of course the EPRB state
\beq
| \Psi \rangle = \frac {1} {\sqrt{2}} \left( | \! \uparrow \rangle \otimes | \! \downarrow \rangle
- | \! \downarrow \rangle \otimes | \! \uparrow \rangle \right),
\eeq
where $ |\!\uparrow \rangle $ denotes spin up in the $z$-direction, but we do not
restrict attention to this choice of state.
Measurements are
made on particle $A$ in the directions characterized by unit vectors ${\bf a}_1$ and
${\bf a}_2$ and on particle $B$ in directions ${\bf a}_3$ and $ {\bf a}_4$.
The probabilities for pairs of such measurements, one on $A$, one on $B$ is of the
form,
\beq
p(s_1, s_3) = \langle \Psi| P_{s_1}^{{\bf a}_1} \otimes P_{s_3}^{{\bf a}_3} | \Psi \rangle
\eeq
plus three similar expressions.
The measurements are described by projection operators of the form
\beq
P_{s}^{\bf a} = \half \left( 1 + s {\bf a} \cdot \sigma \right)
\eeq
where $\sigma_i$ denotes the Pauli spin matrices.

The EPRB state has the property that $ \langle {\bf a} \cdot \sigma \rangle = 0$
for all four direction vectors and this simplifies the analysis considerably since
it means that $B_i = 0 $ in Eq.(\ref{ps}). This is not true for more general states
but it can be arranged by a simple unitary transformation on the initial state.
It is easy to find a unitary transformation operator which carries out independent
rotations on subsystems $A$ and $B$ and this has the effect of performing a rotation
on the average Pauli spin matrices for each subsystem, $ \langle \sigma_i^A \rangle
$ and $\langle \sigma_i^B \rangle $. We may choose the rotation on $A$ so that
$ \langle \sigma_i^A \rangle $ becomes orthogonal to ${\bf a}_1$ and ${\bf a}_2$,
and the rotation on $B$ so that $ \langle \sigma_i^B \rangle $ becomes orthogonal
to ${\bf a}_3$ and ${\bf a}_4$. This then sets all four average spins are zero,
as required. This rotation will in general change the correlation functions. However,
since it acts independently on systems $A$ and $B$ it will not change the degree
of entanglement, so it should not affect whether or not the correlation functions
satisfy the CHSH inequalities.

This argument shows that if analyzing the quantum problem, we can without loss of generality work with a state for which the average spins are zero. However, in the most general case, the probabilities are not of quantum-mechanical form. It is clearly very plausible that the probabilities may be invertibly transformed into a set with zero average spin, but we have not proved this. This will be addressed in more detail elsewhere.

\section{Solution Using a Hidden Variable Model}

We now give a simple proof of Fine's theorem by writing down an explicit local hidden variable model for the probabilities. This model is essentially that given by Peres to illustrate the CHSH inequalities \cite{Per2}.
The model consists of a classical particle
which splits into two with equal and opposite angular momenta, $\pm {\bf J}$, and measurements of
the sign of the angular momentum of each particle are made along directions
characterized by unit vectors
$ \a_1, \a_2$ for one particle and $\a_3, \a_4$ for the other.
We focus on the signs of the variables of the form $ {\bf a} \cdot {\bf J} $
where ${\bf J}$ is assumed to be uniformly distributed. The probability
for all four spins is given by
\bea
p(s_1,s_2,s_3,s_4) = \langle
\left(1 + s_1 \ {\rm sgn} ({\bf a}_1 \cdot {\bf J} )\right)
\left(1 + s_2 \ {\rm sgn} ({\bf a}_2 \cdot {\bf J} )\right)
\nonumber \\
\times \left(1 -s_3 \ {\rm sgn} ({\bf a}_3 \cdot {\bf J} )\right)
\left(1  -s_4 \ {\rm sgn} ({\bf a}_4 \cdot {\bf J} )\right)
\rangle
\label{pJ}
\eea
which is clearly non-negative,
where the average is over ${\bf J}$ with a uniform distribution.

The average spins are zero in this model and the correlation functions are all then of the form
\bea
C_{13} &=&- \langle {\rm sgn} \left( {\bf a}_1 \cdot {\bf J} \right)  \ {\rm sgn}
\left( {\bf a}_3 \cdot {\bf J} \right)\rangle
\nonumber \\
&=& -1 + \frac { 2 \theta_{13} } { \pi }
\eea
where $\theta_{13}$ is the angle between the two vectors \cite{Per2} and lies in the
range $ 0 \le \theta_{13} \le \pi$, and similarly for the other three correlation functions.
Hence the correlation functions in this model
reduce to a simple geometric feature, namely the angle between two vectors.
The CHSH inequalities
take the form
\bea
0 &\le& \theta_{13} + \theta_{23} + \theta_{24} - \theta_{14} \le 2 \pi
\label{t1}
\\
0 &\le& \theta_{13} + \theta_{23} - \theta_{24} + \theta_{14} \le 2 \pi
\label{t2}
\\
0 &\le& \theta_{13} - \theta_{23} + \theta_{24} + \theta_{14} \le 2 \pi
\label{t3}
\\
0 &\le& - \theta_{13} + \theta_{23} + \theta_{24} + \theta_{14} \le 2 \pi
\label{t4}
\eea
These are of course satisfied for any orientation of the four vectors
since there exists a probability Eq.(\ref{pJ}) for this model.
One can also
confirm geometrically that these inequalities hold in this model
by examining all the possible orientations of the four vectors (this is set as an exercise
in Peres' book \cite{Per2}).

However, here, we are interested in the converse to this problem: can we match
the non-negative probability Eq.(\ref{pJ}) to any given set of four marginals
sastisfying the CHSH inequalities?
Or in other words, can we always choose the four vectors
in Eq.(\ref{pJ}) to match any given set of the four angles
satisfying the CHSH inequalities Eqs.(\ref{t1})-(\ref{t4})?
It is not hard to see geometrically that this is indeed possible, thereby providing a proof of sufficiency in Fine's theorem.

In essence the hidden variables model provides a sensible
guess for the underlying probability solving the matching problem and our goal is
to show that it actually does the job. Note that this is not guaranteed --
a particular guess for the probability may have a set of correlation functions which do not explore the full range of possible values satisfying the CHSH inequalities, and indeed we will see an example of this
in Section 6.

\begin{figure}[h]
\begin{center}
\includegraphics[width=5in]{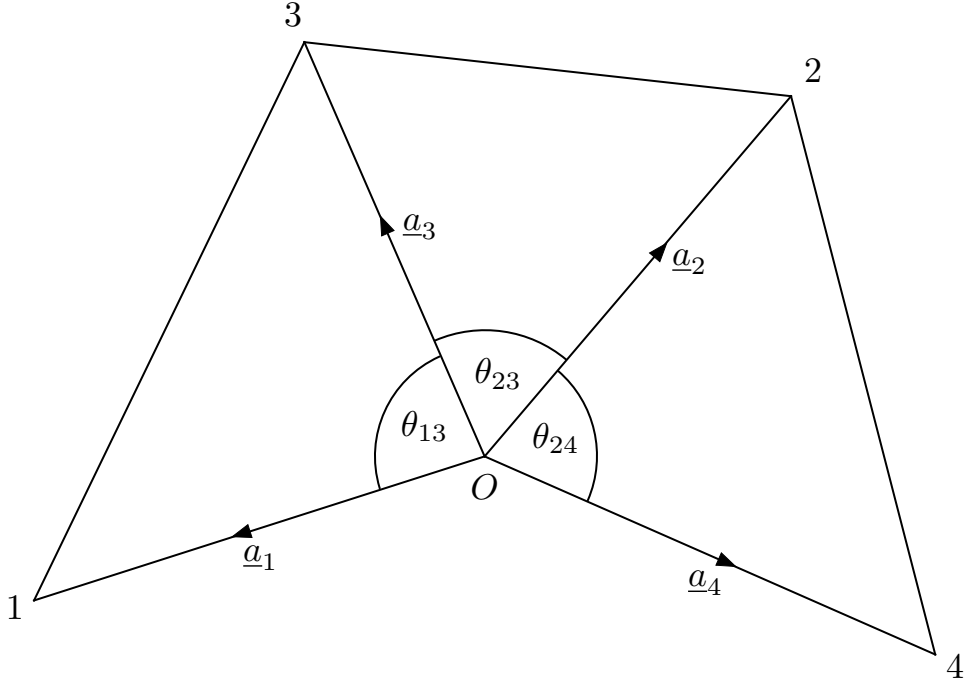}
\caption{A plane figure showing an orientation of the four vectors
$\a_1, \a_2, \a_3, \a_4$ which matches given values for the three angles
$\theta_{13}$, $\theta_{23}$ and $\theta_{24}$. The angle $\theta_{14}$ can be
adjusted, with the first three fixed, by folding in the outer two triangles
along the edges $\a_2$ and $\a_3$, subject to the upper bound Eq.(\ref{upper}).}
\label{fig1}
\end{center}
\end{figure}

We need to show that we can choose the four vectors $\a_1, \a_2, \a_3, \a_4$
to match a given set of angles,
$ \theta_{13}, \theta_{23}, \theta_{24}, \theta_{14} $ satisfying the CHSH inequalities Eqs.(\ref{t1})-(\ref{t4}),
with $\theta_{12}$ and $\theta_{14}$ unspecified. We first let the
four vectors lie in a plane and adjust them so that three of the angles are fixed
to the given values, say $\theta_{13}$, $\theta_{23}$ and $\theta_{24}$. This is shown for a particular orientation of vectors in Figure 1 which shows
three triangles whose edges radiating from the origin $O$ are the four vectors, with
the third side of the triangle completed for illustrative convenience. We then need to adjust these
vectors, by moving them out of the plane, to match the fourth angle $\theta_{14}$, but preserving the three fixed
already. To do this, we imagine that the outer two triangles in the plane figure shown in Figure 1 are allowed to fold inwards along the edges $\a_2$ and $\a_3$, thereby varying the angle $\theta_{14}$
until it reaches its prescribed value. There are limits to the range of values that
can be reached. The largest possible angle is achieved when all four vectors lie
in the plane, as shown in Figure 1, so the upper limit is
\beq
\theta_{14} \le \theta_{13} + \theta_{23} + \theta_{24}
\label{upper}
\eeq
which we know is satisfied since it is one of the CHSH inequalities, Eq.(\ref{t1}).

\begin{figure}[h]
\begin{center}
\includegraphics[width=5in]{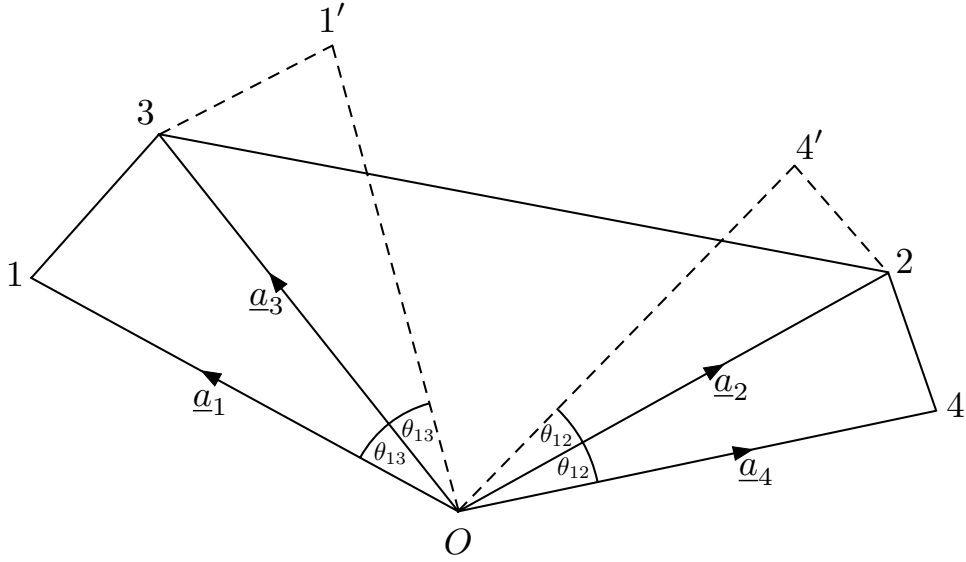}
\caption{
The smallest possible value of $\theta_{14}$, when non-zero, is obtained by folding
in the outer two triangles until they lie flat on the inner triangle, as denoted
by the dotted lines, and thus satisfy the lower bound Eq.(\ref{lower}).
}
\label{fig2}
\end{center}
\end{figure}

A small value of $\theta_{14}$ can be reached by folding in the outer two triangles
as far as possible but it is not always possible to reach $\theta_{14} = 0$ if
one of the three fixed angles is sufficiently large. For example, suppose $\theta_{23}$
is much larger than the other two fixed angles, as depicted in Figure 2. Then the smallest
possible value for $\theta_{14}$ is obtained by folding in the outer two triangles
so they lie flat in the inner triangle, with the vectors $\a_1$ and $\a_4$ lying on
the dotted lines. There is therefore a lower bound on the possible values of $\theta_{14}$
which is easily seen to be
\beq
\theta_{14} \ge \theta_{23} - \theta_{13} - \theta_{24}
\label{lower}
\eeq
which is again seen to be one of the CHSH inequalities, Eq.(\ref{t3}), so will be satisfied.

The orientations depicted in Figures 1 and 2 explore the lower bound on the CHSH inequalities,
Eqs.(\ref{t1})-(\ref{t4}). The upper bound becomes relevant when some of the specified angles are close to $\pi$. One such case is when the unfixed angles $\theta_{12}$ and $\theta_{34}$ are very small and
$\a_1, \a_2$ point in the direction approximately opposite to $\a_3, \a_4$.
In this case, we may work with a different set of vectors in which either the pair
$\a_1, \a_2$ or the pair $\a_3, \a_4$ are reflected in the origin. This has the effect that all four
angles in the CHSH inequalities Eqs.(\ref{t1})-(\ref{t4}) are changed according to
$ \theta \ria \pi - \theta $ and as a consequence the upper and lower bounds are interchanged. Hence we are back to the situation depicted in Figures 1 and 2.

Another case is when three of the vectors, say $\a_1, \a_2, \a_3$, have small angles between them and a large angle with $\a_4$.
In this case, we replace $\a_4$ with its reflection in the origin which causes two of the angles $\theta_{14}$
and $\theta_{24}$ to undergo the transformation $ \theta \ria \pi - \theta $. This creates a more complicated transformation of the CHSH inequalities Eqs.(\ref{t1})-(\ref{t4}) in which again the upper and lower bounds are interchanged, but also some of the CHSH inequalities are interchanged with each other. Again we get back to situations similar to those depicted in Figures 1 and 2.

These arguments can be repeated for other orientations of the vectors and this will
involve the other CHSH inequalities. We thus establish in a simple geometric
way that the parameters of this local hidden variables model, the four vectors, may be chosen in such a way that any set of values of the four correlation functions may be matched, as long as the correlation functions satisfy the CHSH inequalities, and Eq.(\ref{pJ}) is the solution to the matching problem.
This therefore proves sufficiency in Fine's theorem.

\section{The Bell Case}

Turning now to the algebraic proof,
we consider first a simpler example, namely that in which we seek a probability $ p (\s_1, \s_2, \s_3) $
matching the three marginals $ p (\s_1, \s_2)$, $p(\s_2, \s_3)$ and $p(s_1,s_3)$. The probability in this case may be written
\beq
p (\s_1, \s_2, \s_3) = \frac {1} {8} \left( 1 +
\sum_i B_i s_i + \sum_{i < j} C_{ij} s_i s_j + D s_1 s_2 s_3 \right),
\label{qs2}
\eeq
where $i,j,k$ runs over values $1,2,3$. The marginals are obtained by summing out
one of the $s_i$ variables and they have the form Eq.(\ref{p2}). Since these marginals
are, by assumption, non-negative, this imposes certain restrictions on the coefficients
$B_i$ and $C_{ij}$. For example, one obtains a restriction of the form
\beq
1 + B_1 - B_2  - C_{12} \ge 0
\label{res}
\eeq
We are therefore assuming that all such restrictions are satisfied by $B_i$ and $C_{ij}$.
As noted one can argue that $B_i$ may be set to zero, but it is not difficult to maintain
a non-zero value in this proof, so we will do so, thereby seeing explicitly that it
plays essentially no role.

In this example the necessary and sufficient conditions for the existence of a probability are the inequalities \cite{SuZa,Pit}
\bea
C_{12} + C_{13} - C_{23} \le 1
\\
C_{12} - C_{13} + C_{23} \le 1
\\
- C_{12} + C_{13} + C_{23} \le 1
\\
- C_{12} - C_{13} - C_{23} \le 1
\eea
which are a form of Bell's original inequalities \cite{Bell}.
Necessity is easy to establish, along the lines of Eq.(\ref{ss}).
To prove sufficiency, since the three marginals fix the six quantities
$B_i$ and $C_{ij}$, the only free parameter is $D$ so  we need to show that the Bell inequalities ensure that
the constant $D$ can be chosen in such a way that Eq.(\ref{qs2}) is non-negative.

Eq.(\ref{qs2}) is non-negative if
\beq
A(s_1, s_2, s_3) \equiv 1 + \sum_i B_i s_i + \sum_{i < j} C_{ij} s_i s_j  \ge - D s_1 s_2 s_3
\eeq
For the four values of $s_1, s_2, s_3 $ for which $s_1 s_2 s_3 = - 1$, this gives
four upper bounds on $D$,
\beq
A(s_1, s_2, s_3 ) \ge D,
\eeq
and for the values with $s_1 s_2 s_3 = 1$, this give four lower bounds on $D$
\beq
- A(s_1, s_2, s_3) \le D
\eeq
Hence a value of $D$ exists as long as all four upper bounds are greater that the
all four lower bounds:
\bea
&&A(-,-,-),A(+,+,-), A(+,-,+), A(-,+,+)
\nonumber \\
&\ge& - A(+,+,+), -A(-,-,+), -A(-,+,-), -A(+,-,-)
\eea
Of the sixteen resultant inequalities, there are four of the form
\beq
A(s_1,s_2,s_3) + A (-s_1, -s_2, -s_3) \ge 0
\eeq
in which the dependence on $B_i$ drops out and these
are easily seen to be
precisely the Bell inequalities. The remaining twelve are simply the restrictions on $B_i$ and $C_{ij}$ of the form Eq.(\ref{res}) which ensure that the marginals are non-negative.
(This is proved by simply writing them all out, but it is essentially straightforward).
The proves the result.

\section{The CHSH Case}

We now turn to our main algebraic problem, which is proving sufficiency in Fine's
theorem for the CHSH case. We thus seek to show that the CHSH inequalities are a
sufficient condition
for the probability Eq.(\ref{ps}) to be non-negative, for given
values of $B_i$, $C_{13}$, $C_{23}$, $C_{14}$ and $C_{24}$. We use three simplifications.
The first is to restrict attention to the case of zero average spins, $B_i = 0$, as discussed.
The second is to note that the CHSH inequalities are unchanged under the
transformation
\beq
(s_1,s_2,s_3,s_4) \rightarrow (-s_1,-s_2,-s_3,-s_4)
\label{S0}
\eeq
This indicates that a solution to the problem exists which possesses this symmetry.
The probability Eq.(\ref{ps}) with $B_i=0$  will have this symmetry if in addition
$D_{ijk} = 0$, so we make this choice. This means that we will not obtain the most
general solution to the problem, but our aim is to find a reasonably quick way of
showing that a solution exists, which in fact turns out to be not much more complicated
than the Bell case considered in the previous section.
Our third simplification is to
note that the CHSH inequalities have another symmetry, namely
\beq
(s_1,s_2) \rightarrow (-s_1,-s_2)
\label{S1}
\eeq
which, via the symmetry Eq.(\ref{S0}), is equivalent to
\beq
(s_3,s_4) \rightarrow (-s_3,-s_4)
\label{S2}
\eeq
This symmetry is equivalent to changing the signs of the four fixed correlation
functions, $C_{13}$, $C_{14}$, $C_{23}$ and $C_{24}$ whilst preserving the signs
of the unfixed ones $C_{12}$ and $C_{34}$. This symmetry leads to further simplifications
as we shall see.

With the choices $B_i = 0 = D_{ijk}$, the requirement that Eq.(\ref{ps}) is non-negative may be written
\beq
1 + \sum_{i<j} C_{ij} s_i s_j \ge - E s_1 s_2 s_3 s_4
\eeq
which may be written out more explicitly as
\beq
1 + s_1 s_3 C_{13} + s_1 s_4 C_{14} + s_2 s_3 C_{23} + s_2 s_4 C_{24}
\ge - s_1 s_2 C_{12} - s_3 s_4 C_{34} - E s_1 s_2 s_3 s_4
\eeq
We determine the conditions under which these inequalities have a solution.

With the above choices there are only eight inequalities to check, rather than sixteen
and we can select eight independent ones by setting $s_1 = +1$.
Choosing $s_1, s_2, s_3, s_4 $ to be the four sets of values $ (++++) $, $(++--)$, $(+-+-)$,
$ (+--+)$ yields, respectively
\bea
1 + C_{13}+ C_{14} + C_{23} + C_{24}  &\ge &  - C_{12} - C_{34} - E
\label{A1}
\\
1 -  C_{13}- C_{14} - C_{23} - C_{24} &\ge &  - C_{12} - C_{34} - E
\label{A2}
\\
1 + C_{13}- C_{14} - C_{23} + C_{24}  &\ge &  C_{12} + C_{34} - E
\label{A3}
\\
1  - C_{13}+ C_{14} + C_{23} - C_{24}  & \ge &   C_{12} + C_{34} - E
\label{A4}
\eea
Choosing the four values $ (+++-)$, $(++-+)$, $(+-++)$, $(+---)$ yields
\bea
1 + C_{13}- C_{14} + C_{23} - C_{24}  & \ge &  - C_{12} + C_{34} + E
\label{A5}
\\
1 - C_{13}+ C_{14} - C_{23} + C_{24} & \ge & - C_{12} + C_{34} + E
\label{A6}
\\
1 + C_{13}+ C_{14} - C_{23} - C_{24} & \ge &  C_{12} - C_{34} + E
\label{A7}
\\
1  - C_{13}- C_{14} + C_{23} + C_{24} & \ge &   C_{12} - C_{34} + E
\label{A8}
\eea
Now note that the eight inequalities occur
in successive pairs differing only by a reversal of signs of the correlation functions on the left-hand side,
so each pair has the form $ 1 \pm G \ge L $ which are written more concisely as a single relation
$ 1 - | G | \ge L$. This is a consequence of the symmetry Eq.(\ref{S1}).
The eight inequalities therefore reduce to the four inequalities
\bea
1 -  |G_1| &\ge &  - C_{12} - C_{34} - E
\label{G1}
\\
1  - |G_2| & \ge &   C_{12} + C_{34} - E
\label{G2}
\\
1 - | G_3| & \ge & - C_{12} + C_{34} + E
\label{G3}
\\
1  - |G_4| & \ge &   C_{12} - C_{34} + E
\label{G4}
\eea
where
\bea
G_1 &=& C_{13}+ C_{14} + C_{23} + C_{24}
\\
G_2 &=& C_{13}- C_{14} - C_{23} + C_{24}
\\
G_3 &=&  C_{13}- C_{14} + C_{23} - C_{24}
\\
G_4 &=& C_{13}+ C_{14} - C_{23} - C_{24}
\eea
The inequalities Eqs.(\ref{G1})-(\ref{G4}) are now easily solved. Eqs.(\ref{G1}), (\ref{G2})
gives an upper and lower bound on $C_{12} + C_{34}$,
\beq
1 - | G_2 | +E \ \ge \  C_{12} + C_{34} \ \ge \ - 1 + |G_1 | - E
\eeq
which has a solution as long as
\beq
2 - |G_1| - |G_2 | \ge - 2 E
\label{-2E}
\eeq
Similarly, Eqs.(\ref{G3}), (\ref{G4}) yield
\beq
1 - |G_4| - E \ \ge \   C_{12} - C_{34} \ \ge \ - 1 + |G_3| + E
\eeq
which has a solution as long as
\beq
2 - |G_3| - |G_4 | \ge 2 E
\label{2E}
\eeq
Finally, Eqs.(\ref{-2E}), (\ref{2E}) give an upper and lower bound on $E$ and have a solution for $E$ as long as
\beq
|G_1| + |G_2| + |G_3| + |G_4| \le 4
\label{main}
\eeq
This single, simple inequality is a sufficient condition to ensure the non-negativity of
the probability $ p(s_1,s_2,s_3,s_4)$. Written out in full, it reads
\bea
\left| C_{13}+ C_{14} + C_{23} + C_{24} \right| &+&  \left|C_{13}- C_{14} - C_{23} + C_{24}\right|
\nonumber \\
+ \left|C_{13}- C_{14} + C_{23} - C_{24}\right| &+& \left|C_{13}+ C_{14} - C_{23} - C_{24}\right|
\ \le \ 4
\label{main2}
\eea
This is the main result of this section.
Eq.(\ref{main}) or (\ref{main2}) is equivalent to sixteen inequalities corresponding to all the different possible sign choices for $G_1, G_2, G_3, G_4$. It is not immediately obvious but
these sixteen inequalities are, in fact, the eight CHSH inequalities and the eight restrictions of the form $|C_{ij}| \le 1$ on the four fixed correlation functions,
thus proving Fine's theorem.

We briefly outline this last step. We have
\bea
G_1 + G_2 + G_3 + G_4 = 4 C_{13}
\\
G_1 - G_2 - G_3 + G_4 = 4 C_{14}
\\
G_1 - G_2 + G_3 - G_4 = 4 C_{23}
\\
G_1 + G_2 - G_3 - G_4 = 4 C_{23}
\eea
so Eq.(\ref{main}) implies $C_{ij} \le 1 $. The restrictions $C_{ij} \ge - 1$ are
easily found by taking the opposite set of signs for the $G_i's$. Similarly
\bea
G_1 + G_2 + G_3 - G_4 &=& 2 \left(  C_{13} - C_{14} +C_{23} + C_{24} \right)
\\
G_1 + G_2 - G_3 + G_4 &=& 2 \left(  C_{13} + C_{14} - C_{23} + C_{24} \right)
\\
G_1 - G_2 + G_3 - G_4 &=& 2 \left(  C_{13} + C_{14} +C_{23} - C_{24} \right)
\\
G_1 - G_2 - G_3 - G_4 &=& 2 \left( - C_{13} + C_{14} +C_{23} + C_{24} \right)
\eea
Eq.(\ref{main}) then gives upper bound half of the set of CHSH inequalities. The lower bound half is easily obtained by taking the opposite set of signs. (This possibility
is another consequence of the symmetry Eq.(\ref{S1})).

We have therefore shown algebraically that the CHSH inequalities are a sufficient condition for
the non-negativity of the probability Eq.(\ref{ps}), thereby proving Fine's theorem.
The proof hinges on identifying the possibility of setting the average spins to zero
and with making use of the symmetries of the CHSH inequalities. These simplifications
reduce the alegbraic solution of the inequalities on the probabilities to just a few lines.
A side product is an
unusual form of the CHSH inequalities, written as a single inequality, Eq.(\ref{main2}).
This does not appear to have been written down previously
although is closely related to a formula written down in Ref.\cite{ZuBr}. A different but closely related form of the CHSH inequalities was also obtained by Parrott \cite{Par}.

\section{Summary and Conclusions}

In this paper two proofs of Fine's theorem were presented, with the aim to be simple
and pedagogical. The first is based on an explicit local hidden variables model and
the essence of this proof is the simple observation that this model not only satisfies
the CHSH inequalities, as it should, but also provides a complete solution to the CHSH inequalities, in the sense that the parameters of the model may be chosen to
match any values of the correlation functions satisfying the inequalities, hence
the model's probability solves the matching problem.

The second proof is based on a representation of the underlying probability in terms
of correlation functions. This representation highlights a number of simplifying
features, namely the symmetries of the CHSH inequalities. 
The solution obtained for the probability is not the most general
one matching the given marginals, since it involves setting the triple correlator $D_{ijk}$ to zero. General solutions have been given in previous proofs. The essence
of this work is to find the simplest and clearest way to see why the CHSH inequalities
are a sufficient condition for the positivity of the underlying probability matching
the given marginals. A side-product of the investigation is a novel form of the CHSH
inequalities.

In both of these proofs, we assumed that one may set the average spins $B_i$ to zero.
We argued that this is easily achieved in the quantum case, but we have not proved it in general.
This will be addressed in future publications.

Finally, we briefly mention a natural but unsuccessful attempt to prove Fine's theorem,
using a maximum entropy approach \cite{maxent}. As stated, the local hidden variables model used here was essentially a guess as to the form of the probability solving the matching problem. Once in the realm of guessing, it seems reasonable to ask what sort of form for the probability might be the least-biased guess. The maximum entropy method answers this question. The idea is to find the probability which extremizes the entropy
\beq
S = - \sum_{s_1 s_2 s_3 s_4 } p (s_1, s_2, s_3, s_4 ) \ln p (s_1, s_2, s_3, s_4 )
\eeq
subject to the constraints the probability is normalized and the four correlation functions
Eq.(\ref{Cij}) are fixed. We assume the average spins are zero.
The extremization problem is easily solved, with solution
\beq
p (s_1, s_2, s_3, s_4 ) = N \exp \left( \lambda_1 s_1 s_3 + \lambda_2 s_1 s_4 + \lambda_3 s_2 s_3 + \lambda_4 s_2 s_4 \right)
\eeq
where $N$ and the four Lagrange multipliers $\lambda_1, \lambda_2, \lambda_3, \lambda_4$ are to be determined using the normalization condition and expression for the correlation functions Eq.(\ref{Cij}). In effect the maximum entropy method provides a particular ansatz for the solution to
the problem. However, it falls short of solving the problem. At some length, one can show that the algebraic equations for $N$ and the Lagrange multipliers can be solved for sufficiently small $C_{ij}$, but they cannot be solved for the full range of values of the $C_{ij}$ satisfying the CHSH inequalities. For example, when the four $C_{ij}$ are close to $1$, i.e. close to equality
in the CHSH inequalities, there is no solution.

This shows that not every reasonable guess leads to a solution to the problem. Although it may be that a modified version of this problem, perhaps with more quantities fixed, may yield a solution.

\section{Acknowledgements}

I am grateful to Fay Dowker, Terry Rudolph and James Yearsley for useful conversations, and to Stephen Parrott for very useful communications and for pointing out some weaknesses in an earlier draft of this paper.
I also thank James Yearsley for preparing the figures.
This work was supported by EPSRC grant No. EP/J008060/1.

\bibliography{apssamp}

\end{document}